\pdfoutput=1

\documentclass[11pt]{article}

\usepackage[final]{acl}
\usepackage{amsmath}

\usepackage{times}
\usepackage{latexsym}
\usepackage{times}
\usepackage{latexsym}
\usepackage{tabularx}
\usepackage{booktabs}
\usepackage{tabularx}
\usepackage{rotating} 
\usepackage{geometry}
\usepackage{afterpage}
\usepackage{rotating}
\usepackage{tabularx}
\usepackage{multirow}
\usepackage{booktabs}
\usepackage{adjustbox} 
\usepackage{lipsum} 
\usepackage{rotating}

\usepackage{rotating}
\usepackage{booktabs}
\usepackage{tabularx}
\usepackage{graphicx}
\usepackage{amsfonts,amssymb}
\usepackage{booktabs}
\usepackage{array}
\usepackage{cleveref}
\usepackage{paralist}
\usepackage{relsize}

\usepackage[T1]{fontenc}

\usepackage[utf8]{inputenc}

\usepackage{microtype}

\usepackage{inconsolata}

\usepackage{graphicx}
\usepackage{caption}

\title{Language Bias in Information Retrieval: The Nature of the Beast and Mitigation Methods}

  \author{%
  Jinrui Yang\textsuperscript{*}  \qquad
  Fan Jiang\textsuperscript{*} \qquad  
  Timothy Baldwin\textsuperscript{*}\textsuperscript{\dag} \\
  \textsuperscript{*}School of Computing \& Information Systems, The University of Melbourne \\
  \textsuperscript{\dag}Mohamed bin Zayed University of Artificial Intelligence, UAE \\
  {\url{{jinruiy, jifj}@student.unimelb.edu.au}} \qquad  \url{tb@ldwin.net}
}

\begin{document}
\maketitle
\begin{abstract}

Language fairness in multilingual information retrieval (MLIR) systems is crucial for ensuring equitable access to information across diverse languages. This paper sheds light on the issue, based on the assumption that \textit{queries in different languages, but with identical semantics, should yield equivalent ranking lists when retrieving on the same multilingual documents}. We evaluate the degree of fairness using both traditional retrieval methods, and a DPR neural ranker based on mBERT and XLM-R. Additionally, we introduce `LaKDA', a novel loss designed to mitigate language biases in neural MLIR approaches. Our analysis exposes intrinsic language biases in current MLIR technologies, with notable disparities across the retrieval methods, and the effectiveness of LaKDA in enhancing language fairness.
\end{abstract}

\section{Introduction}

Information retrieval (IR) is the process of obtaining relevant information from a large collection of data according to a user's information needs. This information may exist in various formats, including text documents, images, or videos. Conventionally, the collection is a corpus of text documents, and user information needs are expressed in plain text queries. IR serves as a foundational technology in numerous NLP applications including question-answering systems \citep{DBLP:journals/corr/abs-2002-06612, chen-etal-2017-reading}, and is also assuming an increasingly pivotal role in supporting the advancement of Large Language Models (LLMs) for text understanding and knowledge inference \citep{article,zhu2024large}.

Multilingual information retrieval (MLIR) entails queries being in different languages, with the results for a query in a given language being across multiple languages (including the source language of the query). MLIR has particular importance as it enables (multilingual) users to access information that may be unavailable or limited in their native language, thereby fostering cultural and linguistic diversity.

\begin{figure}[t]
    \centering
    \includegraphics[width=0.5\textwidth]{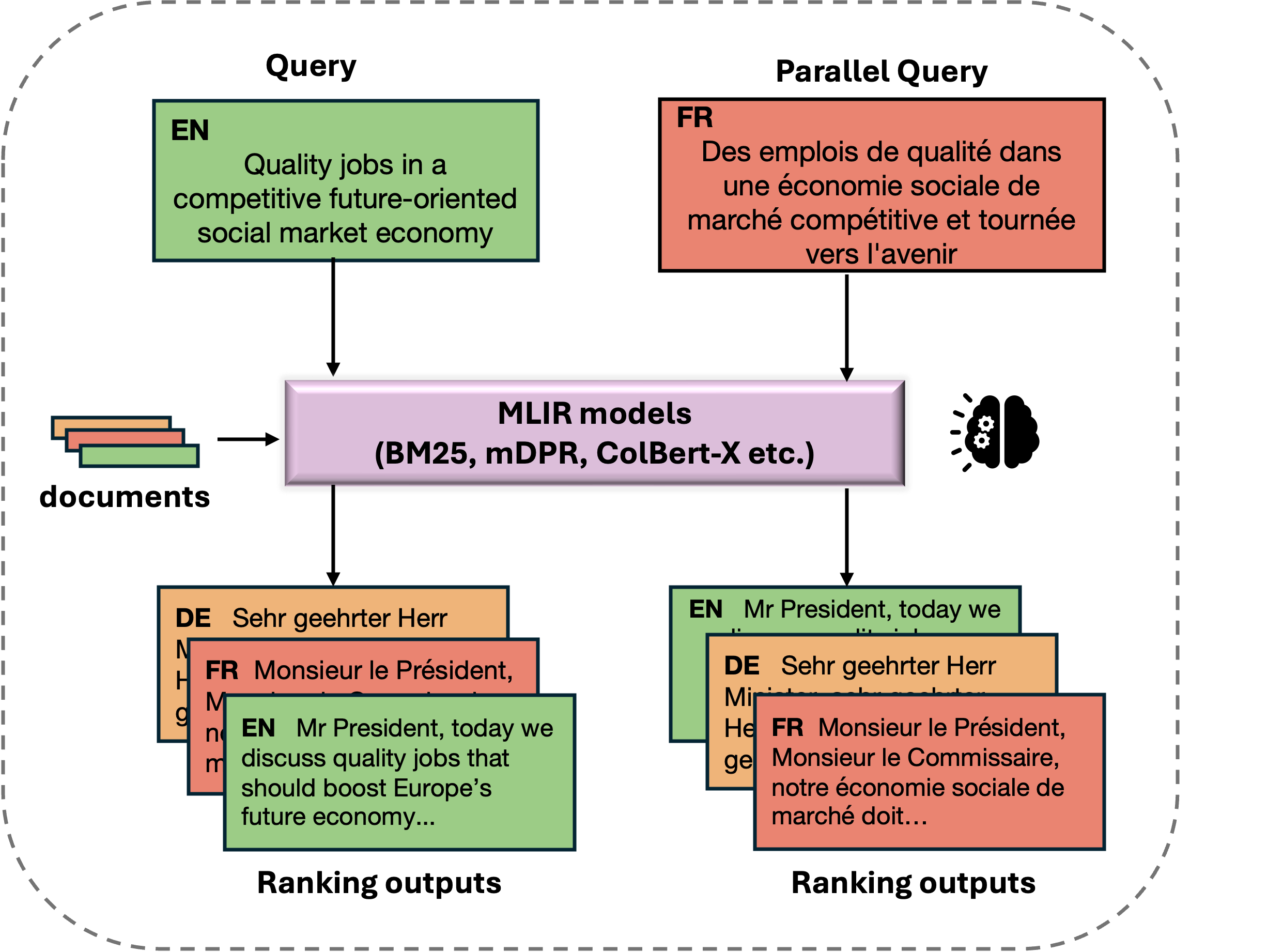}
\caption{The case of language bias studied in this work. Semantically parallel queries retrieve the same documents, but the ranking outputs are inconsistent.}

    \label{introfigure}
\end{figure}

Research has shown that MLIR systems often exhibit biases towards certain languages due to factors like morphological complexity \citep{park-etal-2021-morphology} and resource availability \citep{lawrie2023neural, huang2023soft}. For instance, \citet{lawrie2023neural} found that documents in higher-resource languages tend to be ranked higher in MLIR. This phenomenon is particular notable when the models are built upon multilingual pretrained models \citep{Yang_2024}.

Another case of language bias in MLIR is shown in \Cref{introfigure}. Given semantically equivalent queries in different languages and the same documents, we are interested in determining the consistency of the obtained ranking lists. This forms the basis for our investigation of language fairness in MLIR from a query-level perspective.

Our study compares MLIR methods using semantically equivalent queries in 24 European languages, which we use to search a fixed multilingual document collection. These parallel query sets are from the original dataset, not machine-translated, and are based on human-annotated document tags. In repurposing them as queries and using the tags as relevance judgements, we fashion a multilingual IR collection with massively-multilingual parallel query sets.


Our work makes four main contributions:\footnote{The dataset and code are available from \url{https://github.com/jrnlp/MLIR_language_bias} under an Apache 2.0 license.}
\begin{compactenum}
    \item \textbf{Novel evaluation metric for fairness under ranking:} we propose the mean rank correlation (MRC) score to evaluate language fairness under MLIR, based on the premise that semantically-equivalent queries in different languages should yield consistent document rankings.
    \item \textbf{Novel dataset:} we develop the MultiEup-v2 dataset, consisting of semantically parallel queries and multilingual documents, along with demographic attributes. This dataset serves as a benchmark for future fairness research in MLIR.
    \item \textbf{Quantification of language (un)fairness:} we analyze language fairness in MLIR across different languages and language families, and find that BM25 exhibits larger language bias than neural retrieval frameworks like mDPR. Additionally, higher-resource languages tend to be associated with higher degrees of language fairness.
    \item \textbf{Proposal of a new ranking bias mitigation method:} we propose the language KL-divergence alignment (LaKDA) loss to mitigate language bias in MLIR, demonstrating its effectiveness within the mDPR neural retrieval framework with multilingual text encoders mBERT and XLM-R.

\end{compactenum}

\begin{figure*}[t]
  \includegraphics[width=1.0\linewidth]{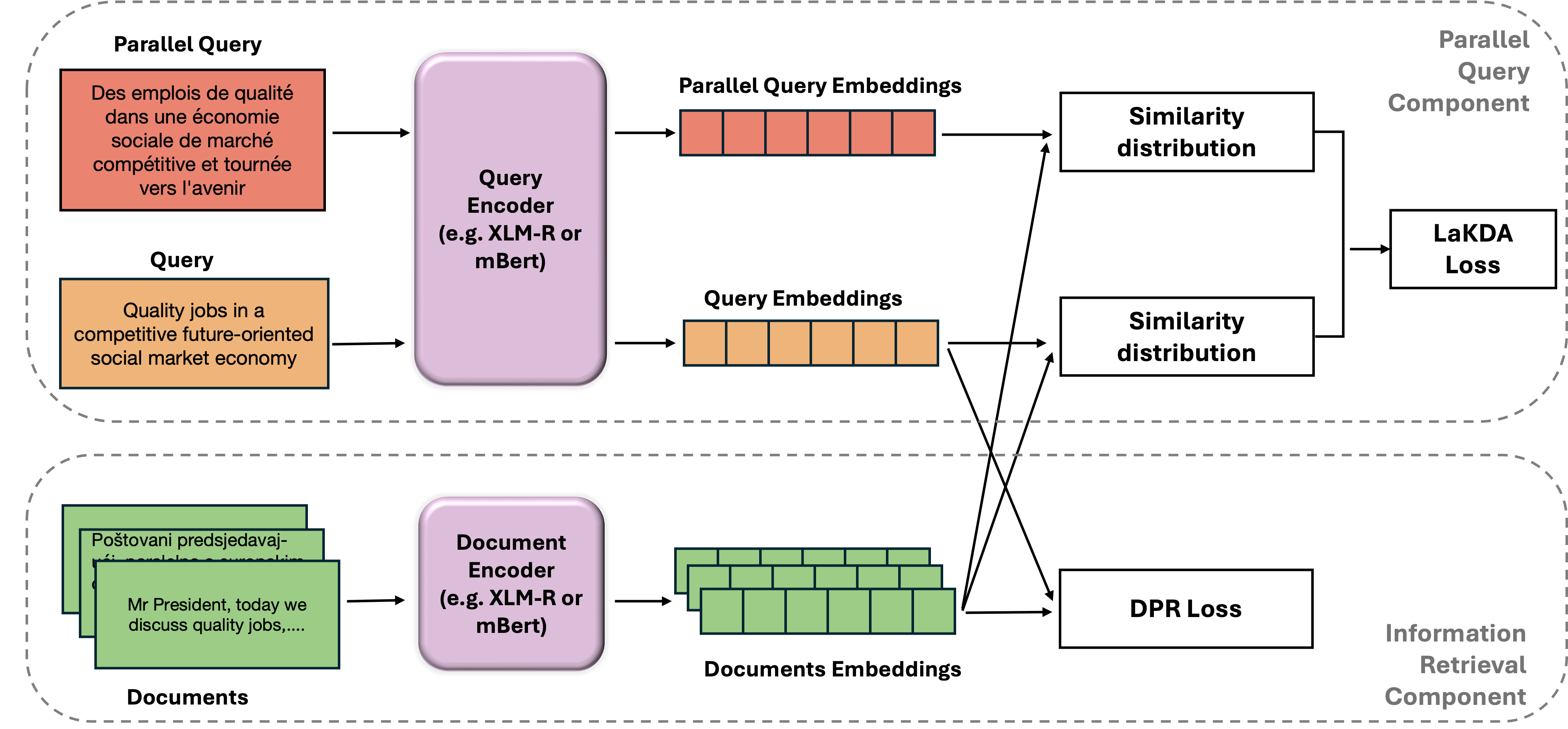} 
  \caption {Our framework contains two parts: the IR component, and the parallel query component. For the IR part, we adopt a DPR module for retrieval with DPR loss. For the parallel query part, we use the LaKDA loss to improve MLIR language fairness. }
  \label{pipeline}
\end{figure*}

\section{Language Bias in MLIR }
In this section, we examine language bias in MLIR. First, we introduce a novel metric for quantifying language fairness, our evaluation benchmark, and introduce a method for mitigating language bias.

\subsection{MLIR Language Fairness Metric} \label{Metric}

We define fairness in MLIR as follows: queries in different languages but with identical semantics should yield equivalent ranking lists when executed against the same multilingual document collection.

Assume we have \( L \) languages and \( N \) queries for each language. For language pair \( a, b \in \{ \ell_1, \ell_2, \ldots, \ell_L \} \), let:
\[
Q_{a} = \{q_{(1,a)}, q_{(2,a)}, \ldots, q_{(N,a)}\}
\]
\[
Q_{b} = \{q_{(1,b)}, q_{(2,b)}, \ldots, q_{(N,b)}\}
\]
represent the sets of all queries in languages \( a \) and \( b \), respectively, where \( q_{(i,a)} \) is the \( i \)-th query in language \( a \) and \( q_{(i,b)} \) is the \( i \)-th semantically parallel query in language \( b \).

Assume a ranking method \( \pi \) produces a ranked result list \( R(q_{(i,a)}, D) \) when given query \( q_{(i,a)} \) against document collection \( D \). 
Then for each query \( i \) and pair of languages \( (a, b) \), we compute the ranking correlation \( \text{RC}_{(a,b)}^i \) between the ranking lists \( R(q_{(i,a)}, D) \) and \( R(q_{(i,b)}, D) \) using Spearman's rank correlation \citep{commentary2010spearman, spearman1904proof}:
\[
\text{RC}_{(a,b)}^i = \text{corr}(R(q_{(i,a)}, D), R(q_{(i,b)}, D))
\]

Next, we compute the average correlation for language \( a\) with query \( i \) with the other \( L-1 \) language pairs, denoted as \( \text{RC}_{(a)}^i\):
\[
\text{RC}_{(a)}^i = \frac{1}{(L-1)} \sum_{1 \leq a < b \leq L} \text{RC}_{(a,b)}^i
\]

The overall mean correlation score (MRC) for a specific language \( a \) among \( L \) languages with \( N \) queries is:
\[
\text{MRC@k}_{(a)} = \frac{1}{N} \sum_{i=1}^{N} \text{RC}_{(a)}^i 
\]
MRC@$k$ represents the average degree of consistency between ranking lists for semantically identical queries across all language pairs in the top-\( k \) results. A higher MRC@$k$ value indicates greater fairness, reflecting a higher degree of equivalence in the search results across different languages.

\subsection{Mitigation Language Bias Methodology } \label{LaKDA}
\Cref{pipeline} demonstrates our co-training MLIR model framework with two losses. \Cref{dprloss} introduces the unitized Dense Passage Retrieval (DPR) loss for IR, and in \Cref{klloss} we propose Language KL-Divergence Alignment (LaKDA) loss to improve language fairness.   
\subsubsection{DPR Loss}  \label{dprloss}
Dense passage retrieval \citep{karpukhin-etal-2020-dense} is a neural retrieval framework initially proposed for monolingual supervised fine-tuning.  This architecture separately encodes queries and documents into dense vectors, optimizing their alignment through a contrastive loss. The goal is to maximize the similarity between queries and their relevant documents while minimizing it with irrelevant documents.

Assume we have a query \( q \) and a collection of documents \( D = \{d^-_1, d^+_2, d^-_3,\ldots, d^-_M\} \), where \( d^+_i \) indicates a relevant document and \( d^-_j \) an irrelevant document.

Let \( \mathbf{q} \) be the dense vector representation of the query, and \( \mathbf{d}^+_i\) and \( \mathbf{d}^-_j \) be dense vector representations of the corresponding documents.
    
The similarity between the query and each document is computed using the dot product: \(\text{sim}(q, d^+_i) = \mathbf{q} \cdot \mathbf{d}^{+\,\intercal}_i\) and \(\text{sim}(q, d^-_j) = \mathbf{q} \cdot \mathbf{d}^{-\,\intercal}_j\).

     We then define the loss to be the negative log-likelihood of the positive documents' similarity scores among all documents:
     \begin{equation*}
       \begin{split}
    \mathcal{L}_{\text{DPR}} = -\frac{1}{N} \sum_{i=1}^{N}
    \log 
    \frac{\exp(\text{sim}(q_i, d^+_i))}{Z_i} \\
  Z_i = \exp(\text{sim}(q_i, d^+_i)) + \sum_{m=1}^{M} \exp(\text{sim}(q_i, d^-_{i,m})).
       \end{split}
     \end{equation*}
This contrastive loss formulation ensures that the query embedding is closer to the positive document embedding than to any of the negative document embeddings, thereby enhancing the model's retrieval performance.


\subsubsection{LaKDA Loss} \label{klloss}
To further mitigate language bias in MLIR, we add a Kullback-Leibler (KL) divergence term to measure the similarity of the distribution of retrieval scores between the original and parallel-language queries over a shared set of document embeddings.

For each query \( q(i, \ell_a) \) and its parallel query \( q(i, \ell_b) \), we compute their similarity distributions over the document embeddings as follows:
\begin{enumerate}

\item \textbf{Compute Similarity Scores:}



For the original query \( q(i, \ell_a) \) and the parallel query \( q(i, \ell_b) \):
\[
\text{sim}(i, \ell) = \mathbf{q}(i, \ell) \cdot \mathbf{D}^\intercal \quad \text{where} \quad \ell \in \{\ell_a, \ell_b\}
\]

\item \textbf{Transform to Probability Distributions:}



The similarity scores are transformed into probability distributions using the softmax function:
\[
\mathbf{p}(i, \ell) = \frac{\exp(\text{sim}(i, \ell))}{\sum_{j=1}^{M} \exp(\text{sim}(i, \ell)_j)}
\]

\item \textbf{KL Divergence Calculation:}

   The KL divergence between the similarity distributions of the original and parallel queries is defined as:
   \begin{eqnarray*}
     \lefteqn{D_{\text{KL}}(\mathbf{p}(i, \ell_b) \parallel \mathbf{p}(i, \ell_a)) =} \\
     && \sum_{j=1}^{M} \mathbf{p}(i, \ell_b)_j \log \left( \frac{\mathbf{p}(i, \ell_b)_j}{\mathbf{p}(i, \ell_a)_j + \epsilon} \right)
   \end{eqnarray*}

   where \(\epsilon\) is a small constant to avoid taking the log of zero.

\item \textbf{Overall LaKDA Loss:}

   The LaKDA Loss for all \( N \) queries is the mean of KL Divergence over all queries:
   \[ \mathcal{L}_{\text{LaKDA}} = \frac{1}{N} \sum_{i=1}^{N} D_{\text{KL}}(\mathbf{p}(i, \ell_b) \parallel \mathbf{p}(i, \ell_a)) \]
\end{enumerate}
Finally, to balance information retrieval performance and language fairness, we define a joint loss function as a weighted combination of the DPR loss \( \mathcal{L}_{\text{DPR}} \) and the LaKDA loss \( \mathcal{L}_{\text{LaKDA}} \):
\begin{equation}
\mathcal{L} = (1-\alpha) \mathcal{L}_{\text{DPR}} + \alpha \mathcal{L}_{\text{LaKDA}}
\label{eq:joint_loss}
\end{equation}
where \( \alpha \) is a tunable hyperparameter.

\subsection{MLIR Language Fairness Benchmark} \label{Benchmark}

\paragraph{Overview} The European Parliament (EP) serves as a crucial forum for political debate and decision-making in the European Union. During debates,
Members of the European Parliament (MEPs) discuss topics in their own languages, and debates are then transcribed in the original languages, and indexed with multilingual topics. 

We constructed MultiEuP-v2 by expanding MultiEuP \citep{yang-etal-2023-multi-eup}, taking the debate titles as queries, and individual MEP speeches in a given debate as documents. The documents are multilingual, encompassing 24 languages from 8 language families. Each query has parallel versions in all 24 languages, sourced from the original dataset.  Additionally, we collected the basic demographic details of each of the MEPs, making it the perfect target for the study of fairness in an IR context, in terms of both language and other protected attributes.

\begin{table}
  \centering
  \begin{tabular}{lcc}
    \hline
    \textbf{} & \textbf{\# Documents} & \textbf{\# Unique Queries} \\
    \hline
    Train           & 44,961                 & 1,623                        \\
    Dev             & 2,787                  & 100                          \\
    Test            & 2,589                  & 100                          \\
    \hline
  \end{tabular}
  \caption{Data size and unique query IDs in train, dev, and test sets. The number of unique query IDs represents the counts for each language.}
  \label{tab:data_size}
\end{table}

\paragraph{Dataset Statistics} We partition the dataset into mutually-exclusive train/dev/test sets to ensure that the queries and documents in the three sets are distinct. \Cref{tab:data_size} details the statistics of the dataset. The number of unique queries is counted per language; i.e., for the dev and test sets, each language has 100 queries, with parallel versions across all 24 languages.\footnote{Note that the selection of languages is defined by those available in the EP data, amongst which there are two low-resource languages: Maltese (MT) and Irish (GA).}  The document collection is also made up of documents from all 24 languages. \Cref{Multi-EuP-stats} in the Appendix shows the language distribution, with languages such as English (EN), German (DE), and French (FR) making up over 50\% of the dataset in terms of document count.

\begin{table*}[t]
\centering
\footnotesize 
\setlength{\tabcolsep}{2pt}
\resizebox{\linewidth}{!}{
\begin{tabular}{lccccccccccccccccccccccccc}
\toprule
\multirow{2}{*}[-1ex]{\textbf{MRR@100}} & \multicolumn{5}{c}{\textbf{Germanic}} & \multicolumn{5}{c}{\textbf{Romance}} & \multicolumn{6}{c}{\textbf{Slavic}} & \multicolumn{3}{c}{\textbf{Uralic}} & \multicolumn{2}{c}{\textbf{Baltic}} & \textbf{Hellenic} & \textbf{Semitic} & \textbf{Celtic} & \\
\cmidrule(lr){2-6}
\cmidrule(lr){7-11}
\cmidrule(lr){12-17}
\cmidrule(lr){18-20}
\cmidrule(lr){21-22}
\cmidrule(lr){23-23}
\cmidrule(lr){24-24}
\cmidrule(lr){25-25}
\textbf{} & \textbf{EN} & \textbf{DE} & \textbf{NL} & \textbf{SV} & \textbf{DA} & \textbf{FR} & \textbf{ES} & \textbf{RO} & \textbf{IT} & \textbf{PT} & \textbf{PL} & \textbf{HR} & \textbf{BG} & \textbf{SK} & \textbf{SL} & \textbf{CS} & \textbf{HU} & \textbf{FI} & \textbf{ET} & \textbf{LT} & \textbf{LV} & \textbf{EL} & \textbf{MT} & \textbf{GA} & \textbf{Avg} \\
\midrule
BM25 & 87.6 & 59.8 & 29.6 & 25.5 & 21.4 & 59.6 & 58.2 & 33.9 & 51.7 & 49.7 & 39.9 & 33.4 & 22.6 & 30.9 & 32.5 & 28.2 & 22.9 & 20.3 & 19.6 & 22.8 & 21.6 & 18.1 & 12.6 & 16.5 & 34.1 \\
\midrule
\multicolumn{25}{l}{\textbf{mBERT} } \\
\midrule
$\mathcal{L}_{\text{DPR}}$  & 42.8 & 39.4 & 37.1 & 36.3 & 33.9 & 38.3 & 41.0 & 38.3 & 39.2 & 40.1 & 39.9 & 37.9 & 36.9 & 39.0 & 39.7 & 37.9 & 32.5 & 32.2 & 30.7 & 35.8 & 33.7 & 30.6 & 27.0 & 14.5 & 35.6 \\
$+\mathcal{L}_{\text{MSE}}$  & 29.1 & 28.7 & 25.8 & 22.5 & 26.5 & 26.2 & 27.0 & 26.2 & 26.9 & 26.8 & 26.9 & 25.9 & 22.8 & 25.7 & 26.0 & 26.3 & 22.7 & 23.7 & 23.3 & 20.0 & 22.9 & 16.3 & 15.6 & 12.2 & 24.0 ($\downarrow$ 32.6\%) \\
$+\mathcal{L}_{\text{LaKDA}}$ & \underline{46.5} & \underline{47.5} & \underline{43.0} & \underline{43.3} & \underline{40.7} & \underline{45.7} & \underline{42.6} & \underline{41.5} & \underline{44.4} & \underline{41.4} & \underline{42.3} & \underline{42.6} & \underline{39.7} & \underline{43.3} & \underline{39.4} & \underline{43.8} & \underline{39.5} & \underline{42.3} & \underline{37.1} & \underline{38.1} & \underline{39.5} & \underline{31.7} & \underline{28.1} & \underline{16.8} & \underline{40.0} ($\uparrow$ \underline{12.4\%}) \\
\midrule
\multicolumn{25}{l}{\textbf{XLM-R} } \\
\midrule
$\mathcal{L}_{\text{DPR}}$  & 47.6 & 49.7 & 45.6 & 49.4 & 45.6 & 48.7 & 51.2 & 51.7 & 45.4 & 47.1 & 47.1 & 45.7 & 51.0 & 50.0 & 43.5 & 49.8 & 43.7 & 47.1 & 44.3 & 46.1 & 49.4 & 46.5 & \textbf{40.6} & 30.0 & 46.5 \\
$+\mathcal{L}_{\text{MSE}}$ & 48.4 & 46.4 & 53.9 & 50.2 & 54.1 & 60.8 & 58.5 & 58.0 & 50.1 & 45.6 & 51.7 & 46.1 & 52.7 & 50.6 & 45.9 & 51.7 & 50.2 & 48.1 & 43.1 & 41.4 & 48.7 & 47.5 & 28.8 & 24.7 & 48.2 ($\uparrow$ 3.7\%) \\
$+\mathcal{L}_{\text{LaKDA}}$ & \textbf{70.0} & \textbf{65.3} & \textbf{65.6} & \textbf{68.8} & \textbf{69.1} & \textbf{69.0} & \textbf{62.0} & \textbf{66.0} & \textbf{67.3} & \textbf{60.7} & \textbf{69.1} & \textbf{57.3} & \textbf{62.2} & \textbf{61.4} & \textbf{64.4} & \textbf{63.4} & \textbf{62.6} & \textbf{59.2} & \textbf{56.7} & \textbf{61.5} & \textbf{60.7} & \textbf{55.4} & 34.9 & \textbf{30.8} & \textbf{61.0} ($\uparrow$ \textbf{31.2\%}) \\
\bottomrule
\end{tabular}
}
\caption{The MLIR performance evaluation results on MultiEuP-v2. MRR@100 ($\times 100$) ranges from 0 to 100, where values closer to 100 indicate better performance. \underline{Underscore} indicates the best performance for mBERT, and \textbf{bold} indicates the best performance for XLM-R. The symbol  \(\uparrow\)/\(\downarrow\) indicates the percentage increase or decrease compared to the vanilla setting \(\mathcal{L}_{DPR}\).  All differences are significant at $p < 0.0005$. Note the broad similarities in results for a given language and also language family.}

\label{mrr}
\end{table*}

\begin{table*}[t]
\centering
\footnotesize 
\setlength{\tabcolsep}{2pt}
\resizebox{\linewidth}{!}{
\begin{tabular}{lccccccccccccccccccccccccc}
\toprule
\multirow{2}{*}[-1ex]{\textbf{MRC@5}} & \multicolumn{5}{c}{\textbf{Germanic}} & \multicolumn{5}{c}{\textbf{Romance}} & \multicolumn{6}{c}{\textbf{Slavic}} & \multicolumn{3}{c}{\textbf{Uralic}} & \multicolumn{2}{c}{\textbf{Baltic}} & \textbf{Hellenic} & \textbf{Semitic} & \textbf{Celtic} & \\
\cmidrule(lr){2-6}
\cmidrule(lr){7-11}
\cmidrule(lr){12-17}
\cmidrule(lr){18-20}
\cmidrule(lr){21-22}
\cmidrule(lr){23-23}
\cmidrule(lr){24-24}
\cmidrule(lr){25-25}
\textbf{} & \textbf{EN} & \textbf{DE} & \textbf{NL} & \textbf{SV} & \textbf{DA} & \textbf{FR} & \textbf{ES} & \textbf{RO} & \textbf{IT} & \textbf{PT} & \textbf{PL} & \textbf{HR} & \textbf{BG} & \textbf{SK} & \textbf{SL} & \textbf{CS} & \textbf{HU} & \textbf{FI} & \textbf{ET} & \textbf{LT} & \textbf{LV} & \textbf{EL} & \textbf{MT} & \textbf{GA} & \textbf{Avg} \\
\midrule
BM25 & 0.5 & $-$1.0 & 1.4 & $-$0.6 & $-$0.5 & $-$1.2 & 0.4 & 2.0 & 2.8 & 1.1 & $-$0.3 & 1.8 & 3.2 & $-$0.3 & 1.5 & 3.3 & 1.7 & 0.4 & 0.6 & 0.7 & $-$2.4 & 1.0 & $-$1.4 & $-$0.5 & 0.6 \\
\midrule
\multicolumn{25}{l}{\textbf{mBERT} } \\
\midrule
$\mathcal{L}_{\text{DPR}}$  & 12.9 & 15.0 & 15.9 & 15.2 & 12.8 & 12.9 & 15.4 & 15.1 & 14.4 & 13.7 & 14.6 & 15.2 & 15.6 & 15.5 & 13.8 & 16.0 & 7.3 & 13.0 & 10.3 & 10.6 & 11.3 & \underline{12.9} & 10.2 & 5.3 & 13.1 \\
$+\mathcal{L}_{\text{MSE}}$ & 14.3 & 15.6 & 12.2 & 14.5 & 14.6 & 15.5 & 13.5 & 18.3 & 17.1 & 15.5 & 15.5 & 14.4 & 10.7 & 15.5 & 14.3 & \underline{17.2} & 13.2 & 14.2 & 12.1 & 12.1 & 12.7 & 6.5 & \underline{13.7} & \underline{7.7} & 13.8 ($\uparrow$ 5.3\%) \\
$+\mathcal{L}_{\text{LaKDA}}$ & \underline{18.3} & \underline{17.4} & \underline{20.1} & \underline{17.4} & \underline{14.9} & \underline{20.6} & \underline{20.2} & \underline{19.1} & \underline{18.2} & \underline{20.7} & \underline{17.3} & \underline{18.2} & \underline{16.9} & \underline{20.0} & \underline{17.5} & 16.7 & \underline{16.9} & \underline{18.7} & \underline{16.0} & \underline{14.2} & \underline{13.0} & 10.2 & 9.4 & 3.5 & \underline{16.5} ($\uparrow$ \underline{25.6\%}) \\ 

\midrule
\multicolumn{25}{l}{\textbf{XLM-R} } \\
\midrule
$\mathcal{L}_{\text{DPR}}$  & 12.6 & 13.8 & 9.3 & 13.9 & 15.2 & 12.7 & 11.7 & 13.1 & 9.5 & 12.7 & 15.0 & 15.7 & 13.0 & 14.1 & 13.7 & 11.2 & 13.8 & 10.0 & 10.9 & 11.3 & 10.6 & 9.5 & 0.2 & \textbf{7.4} & 11.7 \\
$+\mathcal{L}_{\text{MSE}}$ &  12.8 & 11.7 & 11.0 & 12.1 & 13.1 & 9.2 & 12.5 & 12.7 & 13.5 & 11.3 & 11.9 & 12.4 & 11.3 & 9.3 & 11.9 & 10.5 & 11.5 & 10.5 & 9.9 & 11.9 & 9.3 & 8.6 & 2.2 & 0.7 & 10.5 ($\downarrow$ 10.3\%) \\
$+\mathcal{L}_{\text{LaKDA}}$ & \textbf{12.9} & \textbf{20.1} & \textbf{15.8} & \textbf{18.5} & \textbf{18.6} & \textbf{19.3} & \textbf{15.9} & \textbf{16.2} & \textbf{16.9} & \textbf{18.0} & \textbf{17.2} & \textbf{16.0} & \textbf{17.0} & \textbf{16.5} & \textbf{18.3} & \textbf{18.3} & \textbf{17.5} & \textbf{16.6} & \textbf{15.5} & \textbf{16.4} & \textbf{14.3} & \textbf{12.9} & \textbf{6.7} & 6.2 & \textbf{15.9} ($\uparrow$ \textbf{35.9\%}) \\
\bottomrule
\end{tabular}
}
\caption{The MLIR fairness evaluation results on MultiEuP-v2. MRC@5
  ($\times$100) ranges from $-100$ to $100$, where values closer to 100 indicate better fairness. \underline{Underscore} indicates the best fairness for mBERT, and \textbf{bold} indicates the best fairness for XLM-R. }
\label{mrc}
\end{table*}

\section{Experiments and Findings}
Our language fairness experiment consists of two main parts: the detection and comparison of language bias among different ranking methods (\Cref{Language Bias Detection}) and the mitigation of fairness bias (\Cref{mitigation}).

\begin{figure*}[t]
  \includegraphics[width=0.5\linewidth]{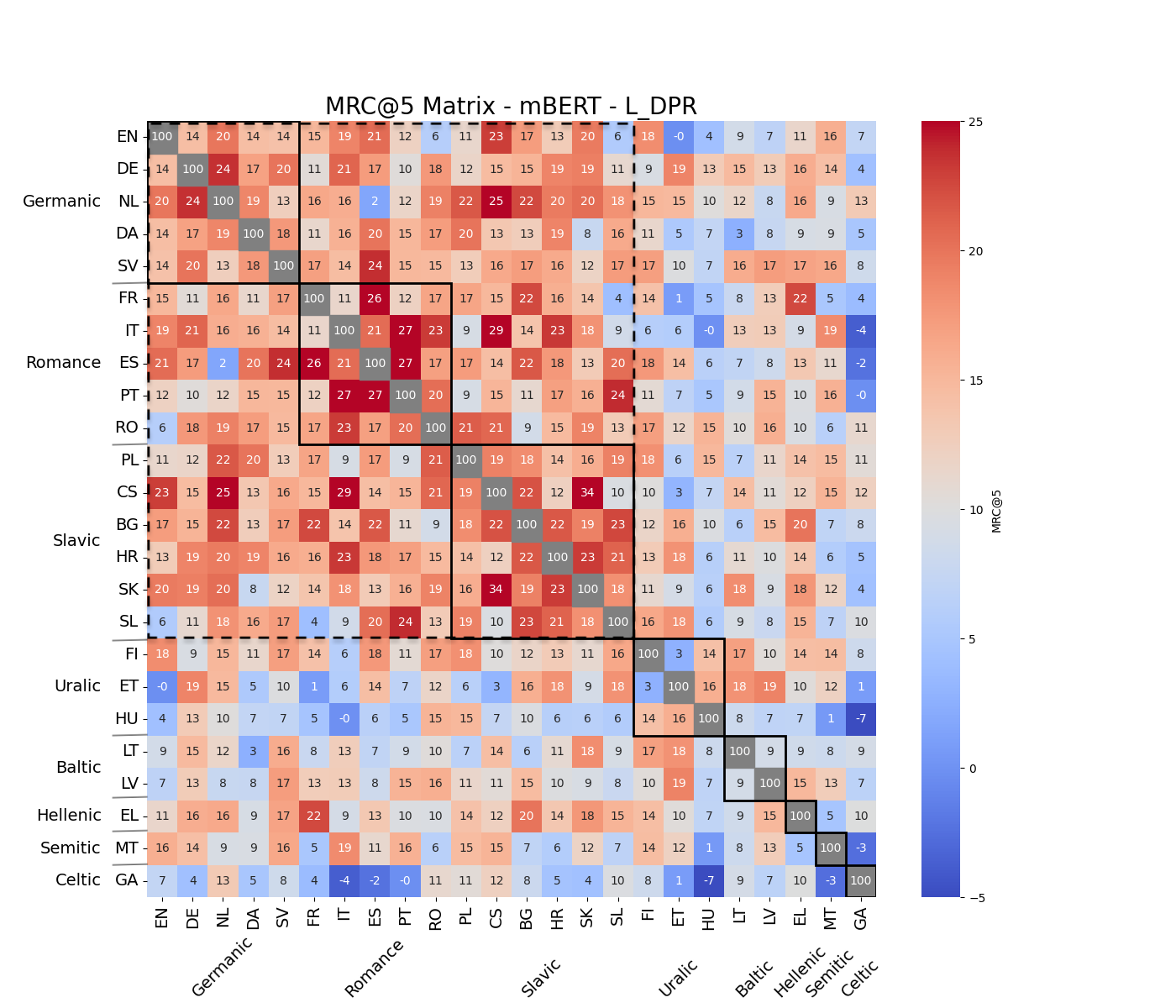} \hfill
  \includegraphics[width=0.5\linewidth]{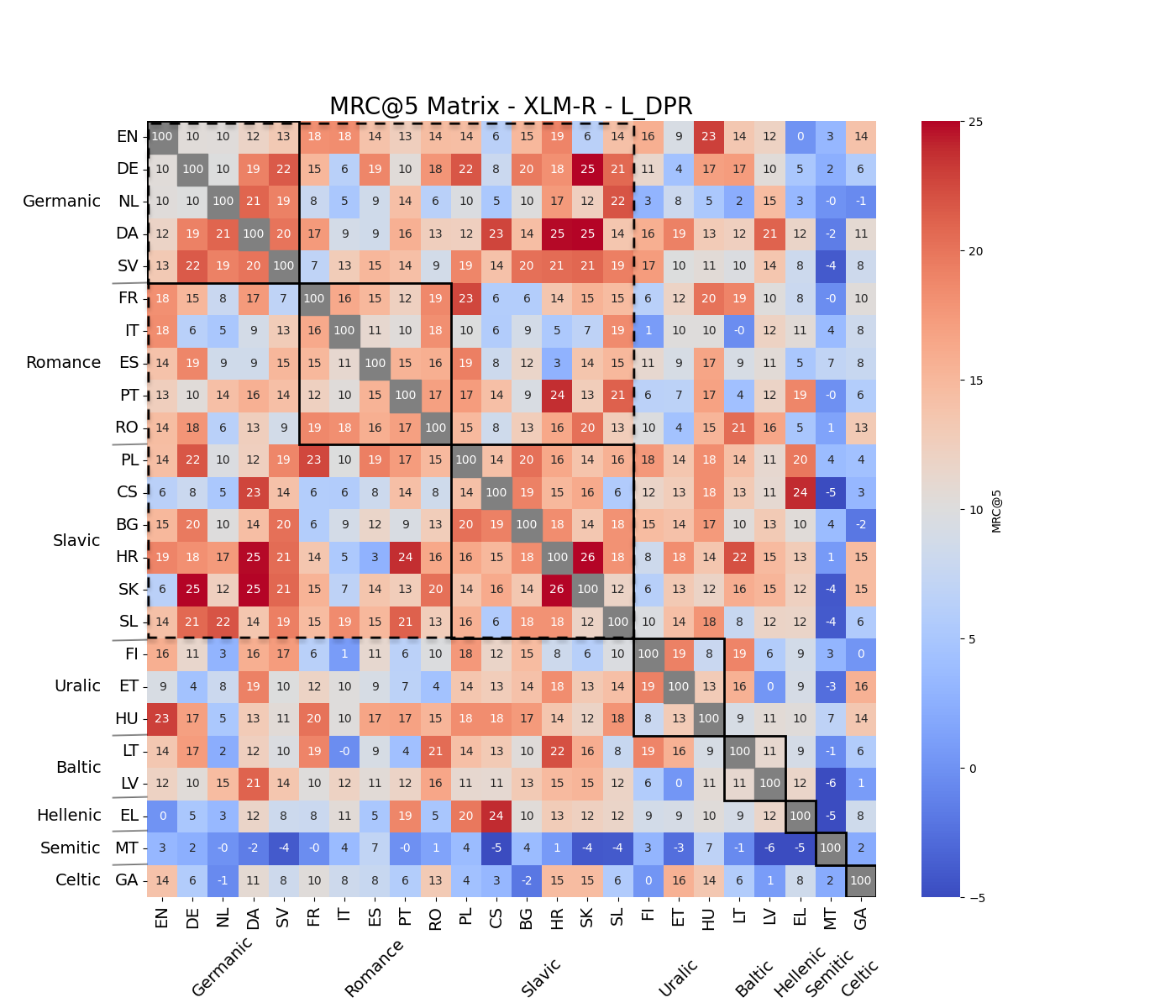}
\caption{The MRC@5 matrix among parallel queries. The x-axis and y-axis both represent query languages.}
  \label{fig:matrix}
\end{figure*}

\subsection{Language Bias Detection} \label{Language Bias Detection}

\subsubsection{Detection Experiment Setting}

We used the MultiEuP-v2 dataset in a \textit{many-vs-many} setting for traning, where both queries and documents are multilingual to ensure language diversity. For evaluation, we adopted a parallel \textit{one-vs-many} approach, with queries in one language and documents in multiple languages, enabling parallel comparison across different languages.

\subsubsection{Detection Experiment Models}
\paragraph{BM25} We implemented BM25, a commonly used traditional information retrieval baseline, using Pyserini \citep{inproceedingslin} which is in turn built upon Lucene \citep{yang+:2017}. We used the default settings ($k_1 = 0.9$ and $b = 0.4$) and language-specific analyzers.

\paragraph{DPR} Our neural IR approach is based on  DPR and uses a bi-directional encoder to encode queries and documents separately. We compare DPR performance over two text encoders: mBERT with \textit{bert-base-multilingual-uncased}, and XLM-R with \textit{xlm-roberta-base}. In each case, the batch size was set to 96 and the learning rate was 5e-5, with each epoch taking approximately 40 minutes on a single Tesla V100 GPU.


\subsubsection{Detection Evaluation and Findings}

\paragraph{Performance Metrics} To evaluate MLIR retrieval performance, we used the MRR@100 metric, which represents the Mean Reciprocal Rank for the top 100 documents \citep{radev-etal-2002-evaluating, voorhees-tice-2000-trec}. For a single query, the Reciprocal Rank (RR) is defined as 
$RR = \frac{1}{\text{rank}}$, where \textit{rank} is the position of the highest-ranked relevant document. If no relevant document is returned, the RR is set to 0. For multiple queries $N$, the MRR is the mean of RRs \citep{yang-etal-2023-multi-eup}.
\[
MRR = \frac{1}{N} \sum_{i=1}^N \frac{1}{\text{rank}_i}
\]

\paragraph{Performance Findings} \Cref{mrr} shows the MRR@100 results for semantically identical queries in different languages. The findings include: (1) for low-resource languages\footnote{Defined as those languages in \citet{conneau-etal-2020-unsupervised} with less than 0.5 GiB in training data. } like Maltese (MT) and Irish (GA), the MRR@100 is lower than high-resource languages; (2) interestingly, despite Maltese having more documents than Estonian (ET) in our dataset (\Cref{Multi-EuP-stats}), the MRR@100 disparity suggests that data augmentation alone does not eliminate the inherent bias in pre-trained IR models against low-resource languages; and (3) DPR with mBERT is slightly better overall than BM25, while DPR with XLM-R significantly outperforms both BM25 and DPR with mBERT.

\paragraph{Fairness Findings} When we evaluate language fairness based on MRC@5 (see \Cref{Metric}), we obtain the results shown in \Cref{mrc}. The main findings are: (1) BM25 has lower language fairness than DPR; (2) similarly to MRR@100, low-resource languages (MT and GA) exhibit lower language fairness than high-resource languages; and (3) according to \Cref{fig:matrix}, which shows the MRC@5 correlation between language pairs (noting that the results are symmetric), languages in the same language family (within the black squares) tend to have higher MRC scores, esp.\ for the Germanic, Romance, and Slavic language families (the dashed square).


\subsection{Language Bias Mitigation} \label{mitigation}

Next we turn to the question of language bias mitigation.

\subsubsection{Mitigation Experiment Setting} The training parameters and evaluation protocol and metrics used to measure language bias mitigation are consistent with those described in \Cref{Language Bias Detection}.

\subsubsection{Mitigation Experiment Models}

\paragraph{Vanilla}
Our vanilla setting is using only the DPR loss for MLIR \citep{karpukhin-etal-2020-dense} and not incorporating any language fairness loss.

\paragraph{MSE}
Another baseline involves calculating the Mean Squared Error (MSE, \citet{hastie2009elements}) between the embeddings of parallel queries to increase their similarity. We employ the same joint MSE loss with DPR loss.

\paragraph{LaKDA} With our proposed LaKDA debiasing method (\Cref{klloss}), for each query, we randomly sample a semantically identical query in a different language and compute the LaKDA loss, which is then jointly optimized with the DPR loss as shown in \Cref{eq:joint_loss}. For both mBERT and XLM-R, we set \(\alpha = 0.5\) for comparability (but return to investigate the hyperparameter sensitivity in \Cref{lambdatuning}).

\subsubsection{Mitigation Evaluation and Findings}

\begin{figure}[t]
  \includegraphics[width=1\linewidth]{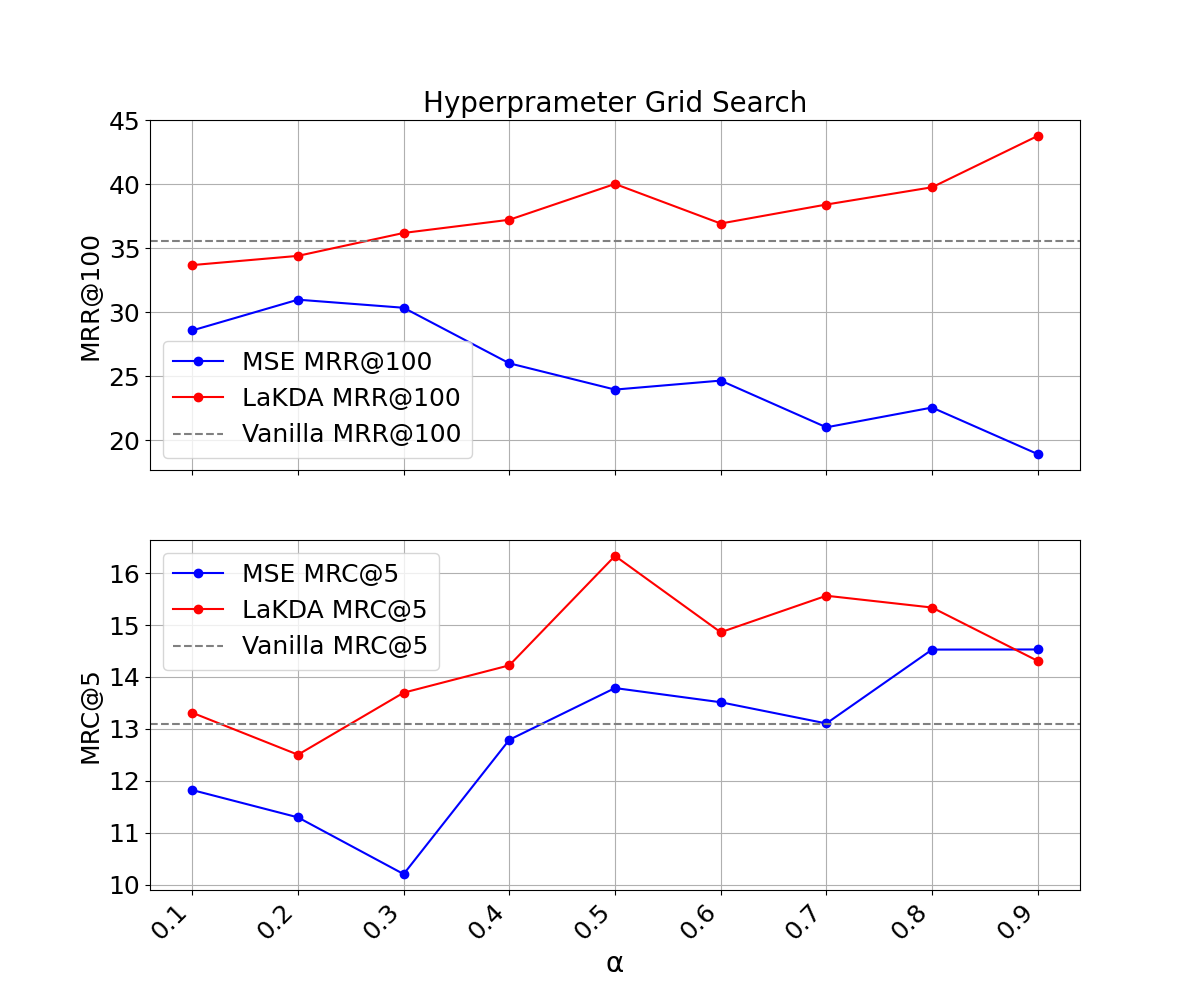} \hfill
  \caption{MSE and LaKDA sensitivity plot. }
  \label{lambdatuning}
\end{figure}

\Cref{mrr} presents the IR performance (MRR@100), and \Cref{mrc} demonstrates language fairness (MRC@5) for the DPR framework with different pretrained multilingual models. Our observations are as follows:

\paragraph{mBERT Findings}
Compared to the vanilla setting (DPR only):
(1) incorporating either MSE or LaKDA enhances language fairness (MRC@5) with mBERT, but LaKDA is more effective (25.6\% vs.\ 5.1\%); and
(2) for MRR@100, LaKDA achieves an 11.3\% improvement, whereas MSE loss reduces MRR@100 by 32.6\%. \Cref{lambdatuning} also shows that during the hyperprameter \(\alpha\) grid search, the DPR model with LaKDA loss is more robust in terms of MRR than MSE loss. This is because, unlike MSE loss, LaKDA loss considers not only the similarity between parallel queries but also their embedding similarity with documents, providing a better trade-off between fairness and performance.

\paragraph{XLM-R Findings}
Compared to the vanilla setting (DPR only):
(1) only LaKDA improves language fairness (MRC@5), by 35.9\%, while MSE leads to a slight degradation; and
(2) both MSE and LaKDA improve IR performance (MRR@100), with increases of 3.7\% and 16.6\%, respectively. XLM-R not only achieves better IR performance but is also more robust. This observation aligns with other research, and is why XLM-R is more commonly used in MLIR \citep{hu2020xtreme,conneau-etal-2020-unsupervised, conneau2019cross}.

\section{Discussion}
\subsection{Improvement of Parallel Query Similarity}
\begin{figure}[t]
    \centering
    \includegraphics[width=0.5\textwidth]{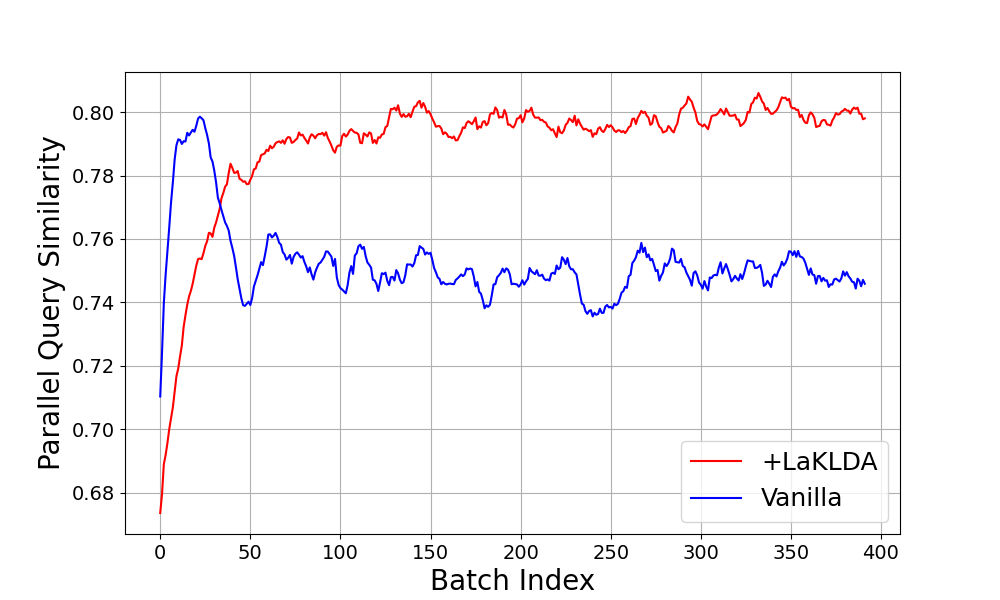}
    \caption{Parallel query similarity over training.}
    \label{fig:query_similarity}
\end{figure}

In our experimental setup, an important characteristic for enhancing language fairness is the increase in similarity of semantically parallel queries. We calculated the average parallel query similarity in each batch over training for mBERT, as depicted in \Cref{fig:query_similarity}. We observe that with the addition of the LaKDA loss, the final stable value of parallel query similarity is higher compared to the vanilla setting. This result explains the enhancement in language fairness (MRC).

\begin{figure*}[t]
  \includegraphics[width=0.45\linewidth]{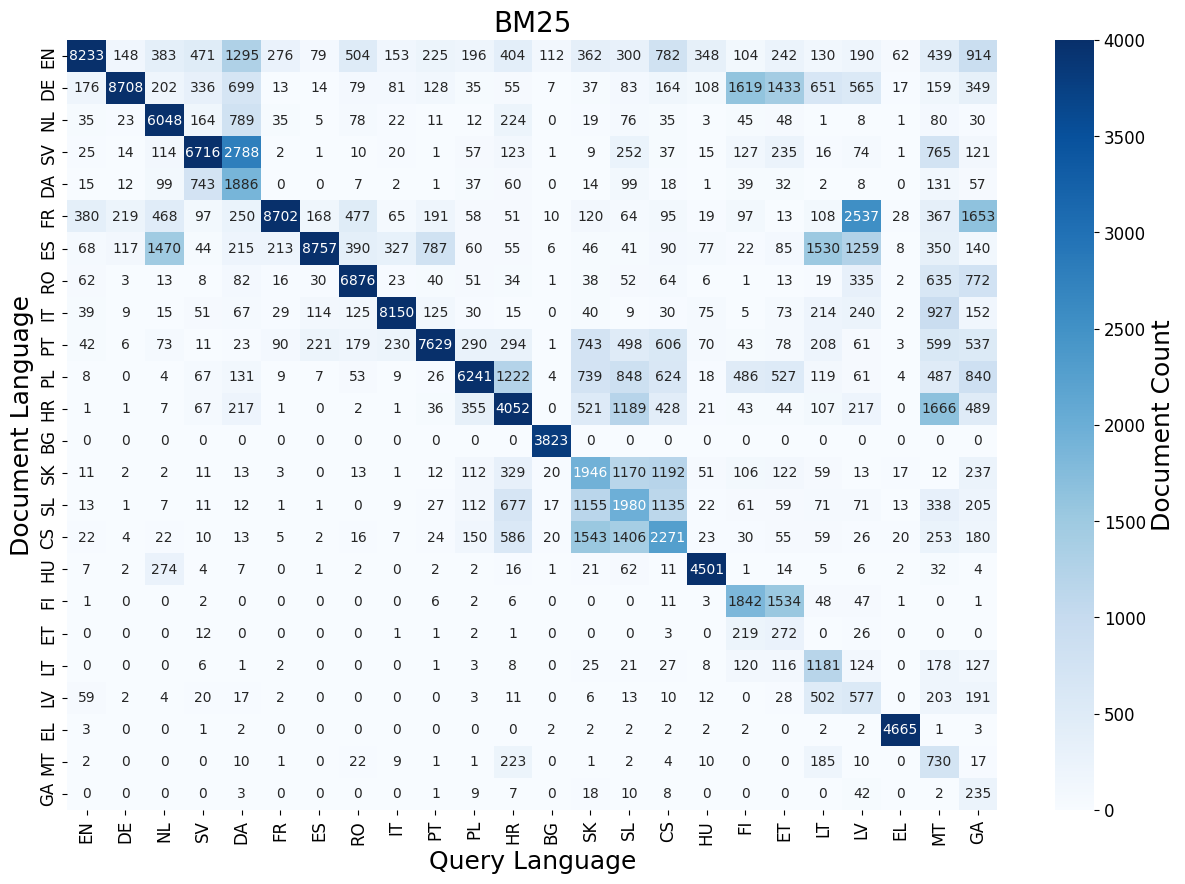} \hfill
  \includegraphics[width=0.45\linewidth]{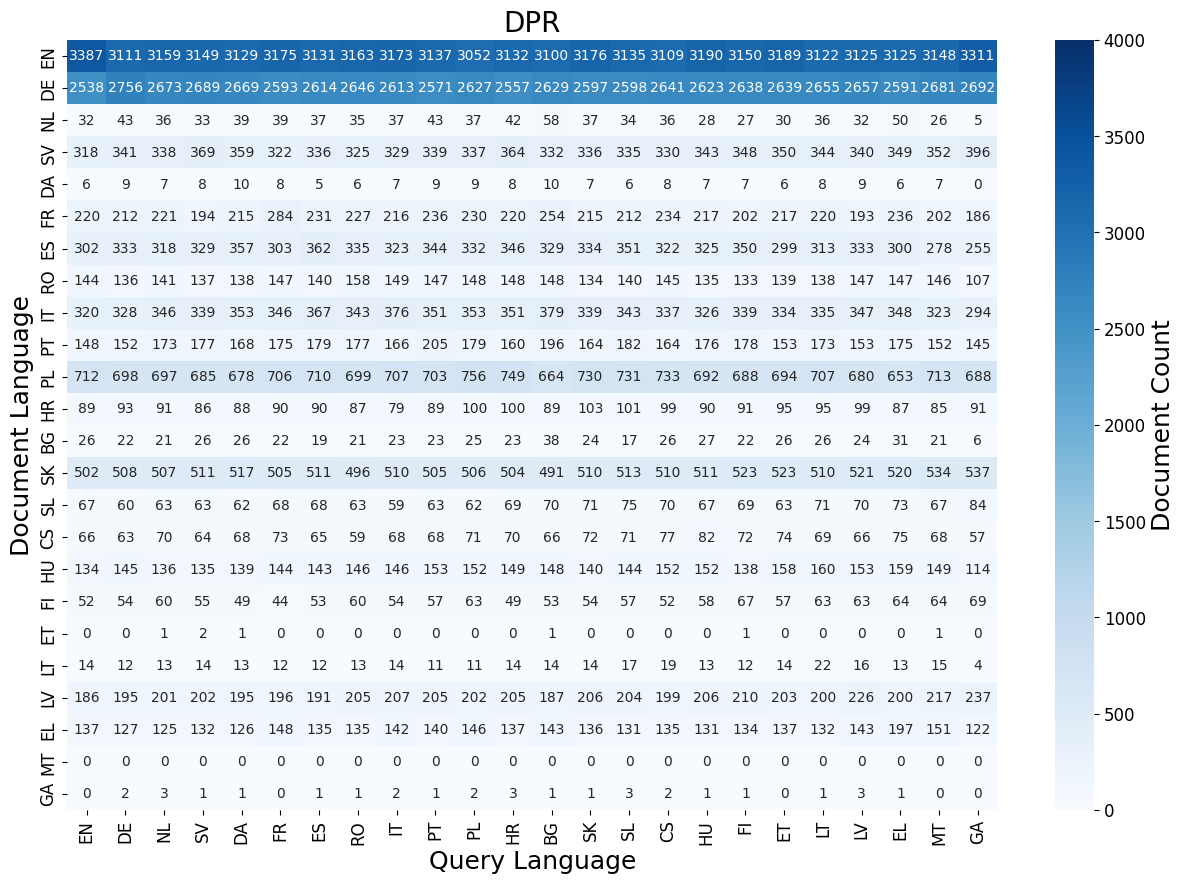}
  \caption {The correlation of query language with document language in top 100 ranking output.}
  \label{fig:correlation}
\end{figure*}
\subsection{Effect of Size and Quality of Parallel Queries}
To explore the impact of the number and quality of parallel queries on IR performance and language fairness, we selected queries in two languages, MT and GA, from the training dataset and conducted experiments under the following three settings:

\textbf{Zero-shot:} As low-resource languages, there is relatively little training data for MT and GA; we therefore excluded queries in MT and GA from the training dataset, keeping the other parallel queries unchanged, and then conducted the same training and evaluation settings.

\textbf{Translation:} Without the original MT and GA parallel queries, we translated English queries into MT and GA parallel queries using Google Translate.\footnote{\url{https://translate.google.com/}} The BLEU scores \citep{papineni2002bleu} of the translation results compared to the original were 0.196 and 0.251, respectively.

\textbf{Original:} The original queries in MT and GA, as per the experiments in \Cref{mitigation}.

\textbf{Findings:} \Cref{mt_query,ga_query} show the results for MT and GA, conducted on the XLM-R model with LaKDA loss. We found that:
\begin{compactenum}
    \item The zeroshot setting had the worst MRR performance, indicating the importance of parallel queries.
    \item The translated version serves as a silver-standard, with improvements in MRR compared to the zeroshot setting.
    \item The original texts are the best choice, achieving the best MRR and MRC, demonstrating the value of our MultiEuP-v2 dataset in providing an original multilingual corpus.
\end{compactenum}
\begin{table}[t]
\small
  \centering
  \setlength{\tabcolsep}{6pt} 
  \begin{tabular}{lcc}
    \midrule
     \textbf{Parallel MT Query} & \textbf{MRR@100}  & \textbf{MRC@5} \\
    \midrule

    Zero-shot    & 21.2  & 2.8 \\
    Translated   & \textbf{36.2}  & 1.2 \\    
    Original    & 34.9  & \textbf{6.7} \\
    \midrule

  \end{tabular}
\caption{Maltese (MT) query MLIR results. }
  \label{mt_query}
\end{table}

\begin{table}[t]
\small
  \centering
  \setlength{\tabcolsep}{6pt} 
  \begin{tabular}{lcc}
    \midrule
     \textbf{Parallel GA Query} & \textbf{MRR@100}  & \textbf{MRC@5} \\
    \midrule

    Zero-shot    & 21.4  & 0.6 \\
    Translated   & 29.6  & 1.4 \\
        Original    & \textbf{30.8}  & \textbf{6.2} \\
    \midrule

  \end{tabular}
\caption{Irish (GA) query MLIR results.}

  \label{ga_query}
\end{table}

\subsection{Effect of Neural Retrieval Approaches}

The MRC@5 results presented in \Cref{mrc} show more than a 20-fold disparity between BM25 and the neural retrieval ranker DPR, with scores of 0.6 and 11.7, respectively. To understand the underlying causes, we analyzed the top 100 ranking outputs from both methods. As shown in \Cref{fig:correlation}, BM25's output document languages and query languages exhibit a strong correlation along the diagonal line, contributing to heightened language bias. Since BM25 is only able to retrieve documents containing keywords present within the query \citep{DBLP:journals/corr/abs-2104-08663} and suffers from lexical gap \citep{10.1145/345508.345576}, resulting in high retrieval rates for documents in the same language as the query. 

Meanwhile, DPR retrieves documents across different languages more effectively, with substantial off-diagonal values and reflecting the skewness of the dataset (see \Cref{Multi-EuP-stats}). This suggests that neural retrieval approaches can mitigate language bias to leveraging multilingual pre-trained models that understand semantic content regardless of the language.

\section{Related Work}
Fairness in Information Retrieval (IR) has been extensively studied across two primary dimensions: individual fairness and group fairness. These frameworks are crucial in ensuring equitable access to information, addressing concerns related to biases in ranking systems.

\textbf{Individual fairness} refers to the principle that similar items (in this case, documents) should be treated similarly \citep{Biega_2018, dwork2011fairnessawareness}. In IR, this means that if two documents are equally relevant to a query, they should receive similar rankings. A violation of individual fairness occurs when two comparable documents are ranked differently due to irrelevant factors, such as their format or metadata. This concept is rooted in the idea of consistency and uniform treatment, ensuring that the system does not unfairly prioritize or penalize specific documents that are otherwise similar in content and relevance.

\textbf{Group fairness}, on the other hand, ensures that predefined groups (such as demographic groups or, in our case, languages) are treated equitably in the ranking process \citep{sapiezynski2019quantifyingimpactuserattention, 10.1016/j.ipm.2021.102707, Zehlike_2017}. The goal is to prevent bias against any group by ensuring that the system does not favor one group over another. In IR, this often translates to ensuring that documents associated with a protected group (e.g., underrepresented languages or communities) are not systematically ranked lower than those associated with unprotected groups. Group fairness frameworks attempt to mitigate historical and societal biases that might seep into the retrieval process, making sure that members of different groups have equitable access to information. In our work, we extend this concept to multilingual IR, treating each language as a group and ensuring that rankings are fair and consistent across languages.

Two key fairness metrics in group fairness that align with our work are Probability of Equal Expected Rank (PEER) and Attention Weighted Ranked Fairness (AWRF). 

\begin{itemize}
    \item \textbf{PEER} \citep{Yang_2024} is designed to ensure equity in ranking by guaranteeing that documents from different languages are treated equally when they are equally relevant. This metric is particularly valuable for multilingual retrieval, as it addresses the risk of language bias, ensuring that a document's rank does not depend on the language of the query if the content is of similar relevance across languages.
    \item \textbf{AWRF} \citep{sapiezynski2019quantifyingimpactuserattention} assesses group exposure by comparing how documents are distributed across ranked positions relative to a predefined target distribution. This metric focuses on ensuring that documents from all languages receive appropriate visibility within the top-ranked results, balancing relevance and fairness in exposure.
\end{itemize}

While these metrics primarily emphasize document-level fairness, our approach uniquely focuses on query-level fairness. In our context, we argue that the retrieval system should provide consistent performance across languages, ensuring that the language of the query does not affect the user’s ability to access relevant information. This promotes inclusivity, ensuring that users from different linguistic backgrounds experience similar outcomes when interacting with the system, ultimately fostering equal access to information.

\section{Background Knowledge}

\paragraph{MultiEuP} The European Parliament (EP) serves as a crucial forum for political debate and decision-making in the European Union. During debates, Members of the European Parliament (MEPs) discuss topics in their own languages, and debates are then transcribed in the original languages, and indexed with multilingual topics. As such, the data is naturally occurring in the 24 official languages of the EU, and expertly transcribed and multilingually annotated. Additionally, we have access to basic demographic details of each of the MEPs, making it the perfect target for the study of fairness in an IR context, in terms of both language and protected attributes was crafted. The EU has published different language versions of all titles, providing semantically identical queries for investigating language fairness in MLIR.

An earlier version of the MultiEuP dataset was published in 2023 covering debates up to October 2022 \citep{yang-etal-2023-multi-eup}. In this work, we have expanded the dataset using the same data collection and preprocessing procedures, to include debates up to 2024. This doubles the total data volume, and provides a sufficient sample size to research neural ranking methods. We additionally augment each document with comprehensive metadata of the author, including gender, nationality, political affiliation, and age, for use in exploring fairness with respect to protected attributes.

Unlike MLIR datasets such as mMARCO \citep{DBLP:journals/corr/abs-2108-13897}, a multilingual version of the MS MARCO \citep{bajaj2016ms}, that relies on machine translation, our benchmark queries and documents are original rather than translated versions. This reduces noise and ensures the linguistic authenticity of the corpus. 

Another commonly used MLIR datasets Mr.\ TyDi \citep{zhang-etal-2021-mr} and  MIRACL \cite{10.1162/tacl_a_00595} , are actually mixed monolingual IR dataset, since they were structured such that queries in different languages are matched only with documents in the same language. This limits the comparability of results across different languages. Our benchmark addresses this limitation by introducing semantically parallel queries across multiple languages, enabling comprehensive analysis of language fairness in MLIR.


\paragraph{DPR} Dense Passage Retrieval (DPR: \citet{karpukhin-etal-2020-dense}) is a neural retrieval framework initially proposed for monolingual supervised fine-tuning. DPR uses dual encoders: one for encoding queries and another for encoding passages (documents), both based on the BERT architecture \citep{devlin-etal-2019-bert}. The primary advantage of DPR over traditional retrieval models like BM25 is its ability to embed both queries and documents into a shared dense vector space, enabling efficient nearest-neighbor search for retrieval. The relevance of a document to a query is determined by the similarity between their embeddings, typically using the dot product as a similarity measure.



In our work, we employ mDPR using mBERT and XLM-R to handle multilingual queries and documents. These models are fine-tuned on parallel query-document pairs from multiple languages, allowing the system to generalize across different languages. The use of mDPR allows us to explore how multilingual language models handle language biases, which often favor high-resource languages over low-resource ones. Furthermore, we investigate the performance of these models on the MultiEuP dataset, assessing their ability to ensure fair and equitable retrieval across 24 languages, thus promoting fairness in multilingual IR.

\section{Conclusion}
We introduced a novel benchmark, MultiEup-v2, for investigating language fairness in multilingual information retrieval (MLIR) systems. Additionally, we proposed the mean rank correlation (MRC) score to assess language fairness in MLIR systems, which ensures that queries in different languages but with the same semantic meaning retrieve similar documents. 
Our findings indicate that the traditional IR method BM25 exhibits larger language biases than DPR with multilingual pretrained language models. Furthermore, we designed the language KL-divergence alignment (LaKDA) loss to mitigate language bias, and found that incorporating LaKDA loss into DPR improves language fairness substantially without sacrificing retrieval performance.

\section*{Ethics Statement}
The dataset contains publicly-available EP data
that does not include personal or sensitive information, with the exception of information
relating to public officeholders, e.g., the names of the active members of the European Parliament, European Council, or other official administration
bodies. The collected data is licensed under the
Creative Commons Attribution 4.0 International
licence.\footnote{\url{https://eur-lex.europa.eu/content/legal-notice/legal-notice.html}}

\section*{Limitations}
Our investigation into language fairness in multilingual information retrieval (MLIR) is limited to European languages in this work. However, our approaches and evaluation methods are adaptable to other languages. Additionally, we focused exclusively on language fairness, leaving other dimensions of fairness in MLIR, such as demographic fairness, unexplored. We encourage the research community to conduct more comprehensive studies on fairness in MLIR, building upon the foundation of our benchmark.

\section*{Acknowledgements}
We sincerely thank Trevor Cohn for his valuable suggestions and support. This research was funded by a Melbourne Research Scholarship to the first author, and was conducted using the LIEF HPC-GPGPU Facility hosted at The University of Melbourne. This facility was established with the assistance of LIEF Grant LE170100200.

\bibliography{custom}

\appendix

\section{Appendix}
\label{sec:appendix}

\begin{table*}[t]
\centering
\renewcommand{\arraystretch}{0.95}
\begin{tabularx}{\linewidth}{Xl}
\begin{tabular}{@{} m{1.7cm}m{0.9cm}m{5.1cm}m{2cm}m{1cm}m{1.2cm}m{1.5cm} @{}}
\toprule
\textbf{Language} & \textbf{ISO code} & \textbf{Countries where official lang.} & \textbf{Language Family} & \textbf{Total Usage} & \textbf{\# Docs} & \textbf{Words per Doc} \\
\midrule
English & EN & United Kingdom, Ireland, Malta & Germanic & 51\% & 14086 & 271/192 \\
German & DE & Germany, Belgium, Luxembourg & Germanic & 32\% & 5861 & 183/168 \\
French & FR & France, Belgium, Luxembourg & Romance & 26\% & 5313 & 267/210 \\
Italian & IT & Italy & Romance & 16\% & 3378 & 191/176 \\
Spanish & ES & Spain & Romance & 15\% & 4621 & 228/195 \\
Polish & PL & Poland & Slavic & 9\% & 2857 & 150/142 \\
Romanian & RO & Romania & Romance & 5\% & 1482 & 183/172 \\
Dutch & NL & Netherlands, Belgium & Germanic & 5\% & 1642 & 180/166 \\
Greek & EL & Greece, Cyprus & Hellenic & 4\% & 1104 & 180/171 \\
Hungarian & HU & Hungary & Uralic & 3\% & 979 & 131/131 \\
Portuguese & PT & Portugal & Romance & 3\% & 2185 & 183/169 \\
Czech & CS & Czech Republic & Slavic & 3\% & 913 & 155/143 \\
Swedish & SV & Sweden & Germanic & 3\% & 1038 & 168/154 \\
Bulgarian & BG & Bulgaria & Slavic & 2\% & 737 & 190/171 \\
Danish & DA & Denmark & Germanic & 1\% & 498 & 206/191 \\
Finnish & FI & Finland & Uralic& 1\% & 564 & 115/111 \\
Slovak & SK & Slovakia & Slavic & 1\% & 698 & 158/157 \\
Lithuanian & LT & Lithuania & Baltic & 1\% & 250 & 145/125 \\
Croatian & HR & Croatia & Slavic & <1\% & 995 & 175/162 \\
Slovene & SL & Slovenia & Slavic & <1\% & 549 & 188/158 \\
Estonian & ET & Estonia & Uralic & <1\% & 88 & 167/162 \\
Latvian & LV & Latvia & Baltic & <1\% & 176 & 128/113 \\
Maltese & MT & Malta & Semitic & <1\% & 243 & 151/148 \\
Irish & GA & Ireland & Celtic& <1\% & 80 & 179/163 \\
\bottomrule
\end{tabular}
\caption{MultiEuP-v2 statistics, broken down by language: ISO language code; EU member states using the language officially; language family; proportion of the EU population speaking the language \citep{chalkidis-etal-2021-multieurlex}; number of debate speech documents; and words per document (mean/median).}
\label{Multi-EuP-stats}
\end{tabularx}
\end{table*}

\begin{table*}[t]
\centering
\footnotesize 
\setlength{\tabcolsep}{2pt}
\resizebox{\linewidth}{!}{
\begin{tabular}{lccccccccccccccccccccccccc}
\toprule
\multirow{2}{*}[-1ex]{\textbf{Recall@100}} & \multicolumn{5}{c}{\textbf{Germanic}} & \multicolumn{5}{c}{\textbf{Romance}} & \multicolumn{6}{c}{\textbf{Slavic}} & \multicolumn{3}{c}{\textbf{Uralic}} & \multicolumn{2}{c}{\textbf{Baltic}} & \textbf{Hellenic} & \textbf{Semitic} & \textbf{Celtic} & \\
\cmidrule(lr){2-6}
\cmidrule(lr){7-11}
\cmidrule(lr){12-17}
\cmidrule(lr){18-20}
\cmidrule(lr){21-22}
\cmidrule(lr){23-23}
\cmidrule(lr){24-24}
\cmidrule(lr){25-25}
\textbf{} & \textbf{EN} & \textbf{DE} & \textbf{NL} & \textbf{SV} & \textbf{DA} & \textbf{FR} & \textbf{ES} & \textbf{RO} & \textbf{IT} & \textbf{PT} & \textbf{PL} & \textbf{HR} & \textbf{BG} & \textbf{SK} & \textbf{SL} & \textbf{CS} & \textbf{HU} & \textbf{FI} & \textbf{ET} & \textbf{LT} & \textbf{LV} & \textbf{EL} & \textbf{MT} & \textbf{GA} & \textbf{Avg} \\
\midrule
BM25 & 77.7 & 75.5 & 77.7 & 75.5 & 75.5 & 68.1 & 75.5 & 76.6 & 77.7 & 76.6 & 74.5 & 77.7 & 75.5 & 76.6 & 75.5 & 74.5 & 75.5 & 74.5 & 75.5 & 77.7 & 76.6 & 76.6 & 75.5 & 62.8 & 75.2
 \\
\midrule
\multicolumn{25}{l}{\textbf{mBERT} } \\
\midrule
$\mathcal{L}_{\text{DPR}}$  & 88.3 & 89.4 & 88.3 & 87.2 & 88.3 & 89.4 & 89.4 & 88.3 & 90.4 & 87.2 & 88.3 & 88.3 & 87.2 & 88.3 & 86.2 & 87.2 & 85.1 & 86.2 & 86.2 & 88.3 & 86.2 & 86.2 & 85.1 & 73.4 & 87.0
\\
$+\mathcal{L}_{\text{MSE}}$ & 74.5 & 72.3 & 72.3 & 71.3 & 72.3 & 66.0 & 72.3 & 72.3 & 73.4 & 73.4 & 71.3 & 73.4 & 71.3 & 72.3 & 71.3 & 69.1 & 70.2 & 72.3 & 72.3 & 69.1 & 70.2 & 71.3 & 67.0 & 60.6 & 70.9
 \\
$+\mathcal{L}_{\text{LaKDA}}$ & 77.7 & 78.7 & 76.6 & 76.6 & 77.7 & 79.8 & 79.8 & 81.9 & 78.7 & 78.7 & 78.7 & 79.8 & 77.7 & 76.6 & 75.5 & 77.7 & 77.7 & 77.7 & 78.7 & 76.6 & 76.6 & 76.6 & 73.4 & 72.3 & 77.6
\\
\midrule
\multicolumn{25}{l}{\textbf{XLM-R} } \\
\midrule
$\mathcal{L}_{\text{DPR}}$  & 86.2 & 89.4 & 86.2 & 84.0 & 86.2 & 85.1 & 89.4 & 89.4 & 86.2 & 87.2 & 86.2 & 84.0 & 87.2 & 88.3 & 90.4 & 86.2 & 89.4 & 90.4 & 84.0 & 80.9 & 86.2 & 86.2 & 88.3 & 81.9 & 86.6
 \\
$+\mathcal{L}_{\text{MSE}}$ & 91.5 & 92.6 & 90.4 & 86.2 & 88.3 & 69.1 & 91.5 & 91.5 & 90.4 & 91.5 & 90.4 & 90.4 & 90.4 & 91.5 & 88.3 & 92.6 & 91.5 & 88.3 & 87.2 & 88.3 & 91.5 & 90.4 & 87.2 & 78.7 & 88.7
 \\
$+\mathcal{L}_{\text{LaKDA}}$ & 93.6 & 96.8 & 93.6 & 93.6 & 93.6 & 67.0 & 94.7 & 94.7 & 92.6 & 96.8 & 95.7 & 96.8 & 95.7 & 94.7 & 92.6 & 92.6 & 94.7 & 92.6 & 95.7 & 93.6 & 93.6 & 92.6 & 89.4 & 75.5 & 92.2
 \\
\bottomrule
\end{tabular}
}
\caption{The MLIR additional evaluation results on MultiEuP-v2. Recall@100
  ($\times$100) ranges from $0$ to $100$, where values closer to 100 indicate better performance. }
\label{mrc}
\end{table*}

\end{document}